\documentclass[%
aip,
pof,reprint,
superscriptaddress
]{revtex4-1}

\usepackage{graphicx}
\usepackage{dcolumn}
\usepackage{bm}

\begin{document}






\title{A Comparative Study of an Asymptotic Preserving Scheme and Unified Gas-kinetic Scheme in Continuum Flow Limit}
\author{Songze Chen}
\affiliation{
Hong Kong University of Science and technology \\
Clear Water Bay, Kowloon, Hong Kong, China
}%
\author{Kun Xu}%
\email{makxu@ust.hk}
\affiliation{
Hong Kong University of Science and technology \\
Clear Water Bay, Kowloon, Hong Kong, China
}%

\date{\today}

\begin{abstract}

Asymptotic preserving (AP) schemes are targeting to simulate both continuum and rarefied flows.
Many AP schemes have been developed and are capable of capturing the Euler limit in the continuum regime.
However, to get accurate Navier-Stokes solutions is still challenging for many AP schemes.
In order to distinguish the numerical effects of different AP schemes on the simulation results in the continuum flow limit,
an implicit-explicit (IMEX) AP scheme and the unified gas kinetic scheme (UGKS) based on Bhatnagar-Gross-Krook (BGk) kinetic equation
will be applied in the flow simulation in both transition and continuum flow regimes.
As a benchmark test case, the lid-driven cavity flow is used for the comparison of these two AP schemes.
The numerical results show that the UGKS captures the viscous solution accurately. The velocity profiles are very close to the classical benchmark solutions.
However, the IMEX AP scheme seems have difficulty to get these solutions.
Based on the analysis and the numerical experiments,
it is realized that the dissipation of AP schemes in continuum limit is closely  related to the numerical treatment of collision and transport
of the kinetic equation.
Numerically it becomes necessary to
couple the convection and collision terms in both flux evaluation at a cell interface and the collision source term treatment inside each control volume.

\end{abstract}

\maketitle


\section{Introduction}

The description of  a gas flow depends on the resolution to identify it.
On the macroscopic level, the well-defined governing equations for fluid system are the Euler equations and Navier-Stokes (NS) equations.
On the other hand, if the resolution is fine enough, the Boltzmann equation can be used to describe the flow system microscopically.
The connection between the molecular and macroscopic descriptions of the flow motion has been investigated all the time.
In early 20th century, Chapman and Enskog \cite{Chapman1970} independently deduced the Navier-Stokes equations from the Boltzmann equation under certain asymptotic limits.
The successful achievement in this derivation is that the viscosity and heat conduction coefficients in the Navier-Stokes equations are quantitatively
related to the intermolecular forces.
In particular, this connection is only valid when the Knudsen number ($\mbox{Kn}$), which is defined as the ratio between the mean free path ($\lambda$) and the characteristic
length of the problem, is small, say, $\mbox{Kn} \ll 1$.
Therefore, there certainly exist some circumstances under which the Navier-Stokes equations are inadequate to describe the flow motion.
Traditionally, the Navier-Stokes equations are hard to capture the rarefied flow phenomena as the Knudsen number is not small.
However, in the continuum flow regime, the kinetic theory provides too many detailed information which is not needed.
Therefore, many studies have been focused on the development of a unified numerical scheme,
which simulates different flow regimes efficiently with different approaches, such as kinetic one in the rarefied regime and NS one in the
continuum regime. This paper concerns the kinetic schemes with such a property.
For other possible approaches, such as domain decomposition strategies and hybrid methods, please refer to \cite{Degond2005,Kolobov2007,Tiwari1998}. 

A numerical scheme, which is capable of capturing the characteristic behavior on different scales with a fixed discretization in both space and time,
is called asymptotic preserving scheme \cite{Jin2012}.
Specifically, for a gas system, it requires the AP scheme to recover the NS limit and Euler limit on fixed time step and mesh size
as the Knudsen number goes to zero. A standard explicit scheme for kinetic equation always requires the space and time discretizations
to resolve the smallest scale of the system, such as the particle mean free path.
It makes the scheme tremendously expensive when the system is close to the continuum limit.
In recent years, many studies have been concentrated on the development of AP schemes.
It has been shown that  delicate time \cite{Caflisch1997,Gabetta1997} and space \cite{Jin1996} discretizations should be adopted in order to achieve AP property.
Physically, the continuum limit is achieved by efficient and intensive particle collisions.
The local velocity distribution function is rapidly approaching to a local equilibrium state.
Based on this fact, it is clear that any plausible approximation to the collision process must project the non-equilibrium state to the local equilibrium one.
Previous results \cite{Jin1995b,Caflisch1997,Gabetta1997} show that an effective condition for the numerical scheme to recover the correct continuum limit
is to project the distribution function to the local equilibrium one with a discrete analogue of the asymptotic expansion of the continuous equations.
In these studies, implicit time discretization is implemented to meet requirement of numerical stability and AP property.
On the other hand, it seems that the space discretization is not as crucial as the time discretization in the sense of AP property \cite{Jin2012}.
However, the sophisticated space discretization is still necessary for the capturing of correct long-time behavior of hyperbolic systems with stiff relaxation terms \cite{Jin1996}.
The asymptotic balance must be reflected in the space discretization. Similar principle for space discretization has been proposed
in the gas kinetic scheme (GKS) \cite{Xu93phd,Xu2001}. This scheme doesn't solve the distribution function directly, but uses the conservative variables to reconstruct the initial distribution function with Chapman-Enskog expansion. The essential ingredient of GKS is the implementation of a local analytical solution
which allows the collision effect to participate in construction of the space discretization of the equilibrium state.

Although all AP schemes are able to obtain the Euler limit, only a few studies target to the AP property in the sense of recovering NS solutions \cite{Bennoune2008,Filbet2010,Filbet2011,ugks1}.
Any dissipation in the AP schemes can be considered as an artificial one in the Euler limit,
as long as it is consistent with the numerical scheme of the Euler equations.
The Navier-Stokes limit is characterized by the dissipation which is well-defined.
The Euler limit can be easily achieved once the scheme is conservative.
However, for the Navier-Stokes limit the numerical dissipation has to be controlled and be less than the physical one.
The operator splitting strategy of decoupling transport and collision always results in the Euler AP property only \cite{Bennoune2008}.
A Navier-Stokes AP property can be obtained by including the convection terms in the collision process \cite{Bennoune2008,Filbet2010,Filbet2011}.
But this kind of schemes involve artificial dissipation which depends on spatial derivatives of the last time step.
As a result, the time accuracy can only have first order.
Bennoune \emph{et. al.} constructed a NS AP scheme using equilibrium and non-equilibrium splitting technique.
Their analysis shows that the dissipation term is identical with the Chapman-Enskog expansion \cite{Bennoune2008}.
However, they ignored the large numerical dissipation in the Euler flux which is evaluated by kinetic flux vector splitting (KFVS) method.

In the past years, the AP schemes and unified gas-kinetic scheme have been developed.
The implicit schemes with the implementation of the convection term into the collision process
is accepted as a strategy for the Euler AP property.
But the AP property for the Navier-Stokes limit has not been extensively tested.
Moreover, a practical and useful multiscale computation is to get a correct transition from kinetic scale to the Navier-Stokes limit, not the Euler limit.
Therefore, the NS AP property has strong engineering application value.
One of the main purpose of this paper is to set up a benchmark test case for the validating of NS limit for any scheme which is targeting to the science and engineering applications.
The Bhatnagar-Gross-Krook (BGK) model is adopted for the development of AP schemes because it is the simplest kinetic model for flow system.
We are going to compare the numerical results of these AP schemes in the lid-driven cavity flow simulations with well-defined reference data \cite{Ghia1982}.
Then, the principle to achieve the NS AP limit will be presented.
In the following, section 2 introduces the AP schemes. Section 3 presents the numerical results.
Section 4 discusses the dissipative mechanism and the relation to the particle collision and convection. Section 5 gives the conclusion.

\section{AP schemes}

\label{sec:APschemes}
\subsection{Bhatnagar-Gross-Krook model}
\label{sec:BGK}
The Bhatnagar-Gross-Krook model was proposed to simplify the Boltzmann equation \cite{Bhatnagar1954}. It takes the form as following,
\begin{equation}
    \frac{\partial f}{\partial t}+\mathbf{u}\cdot\frac{\partial f}{\partial \mathbf{x}} =
    \frac{g-f}{\tau},
    \label{eq:BGKModel}
\end{equation}
where $f(\mathbf{x},\mathbf{u},t)$ represents the particle velocity distribution function. It is a function of location $\mathbf{x}$, particle velocity $\mathbf{u}$, and time $t$. And $\tau$ denotes the relaxation time.
The conservative variables $W$ can be obtained by taking moments of $f$,
\begin{eqnarray}
  W = \left(\begin{array}{l} \rho \\ \rho \mathbf{U} \\ \rho E \end{array}\right) &=& \int \psi f d\mathbf{u} = <f>  , \nonumber
\end{eqnarray}
where $\psi = (1,\mathbf{u}, \frac{1}{2}\mathbf{u}^2)^T$. In this study, we only consider monatomic gas, thus ignore the internal degree of freedom.
The right hand side of Eq.(\ref{eq:BGKModel}) presents the relaxation process. Here $g$ is the equilibrium state which reads,
\begin{equation}
    g = \rho\left(\frac{1}{2\pi RT}\right)^{\frac{3}{2}}e^{-\frac{1}{2RT}(\mathbf{u-U})^2},
    \label{eq:MaxwellState}
\end{equation}
where $T$ denotes temperature and $R$ is gas constant.
It is straightforward to get the Navier-Stokes equations by the Chapman-Enskog expansion of the BGK model.
Since the equilibrium state $g$ is independent of small parameter, say, Knudsen number or intuitively the relaxation time $\tau$,
only the particle distribution function $f$ is needed to expand in a power series in terms of the small parameter $\tau$ when $\tau\rightarrow 0$.
For simplicity, only the first two terms are considered here,
\begin{equation}
f = f^{(0)}+\tau f^{(1)}+O(\tau^2). \label{eq:fExpansion}
\end{equation}
All deviations from the equilibrium state have zero moments for the  conservative variables, namely,
\begin{equation}
 \int \psi f^{(1)} d\mathbf{u} = 0. \label{eq:fConstraint}
\end{equation}
Substituting Eq.(\ref{eq:fExpansion}) into the BGK model, the first order Chapman-Enskog equation for $f^{(0)}$ is obtained
by equating terms of $O(\frac{1}{\tau})$, that is
$$g - f^{(0)} = 0.$$
Up to this order $f = f^{(0)} = g$, the conservative moments of BGK model gives the Euler equations.
The equation corresponding on the order $O(\tau^0)$ is
$$f^{(0)}_t + \mathbf{u}\cdot\nabla f^{(0)} = -f^{(1)},$$
from which  we get,
\begin{eqnarray}
f^{(1)} = -(g_t + \mathbf{u}\cdot\nabla g)  . \label{eq:CEOp1}
\end{eqnarray}
Now the expansion (\ref{eq:fExpansion}) turns out to be,
\begin{equation}
f = g-\tau (g_t + \mathbf{u}\cdot\nabla g)+O(\tau^2). \label{eq:fExpansion1}
\end{equation}
By substituting the formula (\ref{eq:fExpansion1}) into the BGK equation, with the
neglection of higher order terms, we have
\begin{eqnarray} \label{eq:NS}
(f^{(0)}_t+\mathbf{u}\cdot\nabla f^{(0)})+\tau(f^{(1)}_t + \mathbf{u}\cdot\nabla f^{(1)}) = \frac{g-f}{\tau} .
\end{eqnarray}
Taking conservative moments on the above equation gives the Navier-Stokes equations.
Specifically, the moments of the right hand side is zero because the conservative moments of $g$ and $f$ are the same.
The first part of the left hand side is identical with the Euler equations.
Only the second term on the left hand side is related to the shear stress and the heat flux which characterize the Navier-Stokes equations.
The moments of the forepart, $\int \psi\tau f^{(1)}_t d\mathbf{u}$, is $0$ due to the expansion constraint (Eq. (\ref{eq:fConstraint})).
By substituting the formula (\ref{eq:CEOp1}), the remaining part of the conservative moments reads
\begin{equation}
\nabla\cdot(\int\tau\mathbf{u}\psi(-(g_t + \mathbf{u}\cdot\nabla g))) d\mathbf{u}.
\end{equation}
which represents the dissipation terms.
It is a convenient formulation to check the dissipation of an asymptotic model.
As shown by the above formula, the deviation from the equilibrium state, namely, $- \tau (g_t + \mathbf{u}\cdot\nabla g)$, dominate the asymptotic process
and give the Navier-Stokes limit with the inclusion of shear stress and heat conduction terms.
We regard formula (\ref{eq:fExpansion1}) as a rule to measure the dissipation of numerical schemes in the NS limit.

\subsection{Implicit-Explicit AP scheme}
\label{sec:IntrIMEX}
Owing to the stiffness of the BGK relaxation model, the standard explicit scheme requests very small time step.
To overcome this difficulty, implicit scheme has to be adopted when the system behaves asymptotically.
In fact, the implicit term can be solved explicitly thanks to the conservation laws.
Here, we introduce a simple one as illustration \cite{Jin1995b,Filbet2011},
\begin{eqnarray}
\frac{f^{n+1}-f^{n}}{\Delta t}+\mathbf{u}\cdot \nabla f^n = \frac{g^{n+1}-f^{n+1}}{\tau^{n+1}}.
\end{eqnarray}
The convection term will be discretized later.
Here, it is formally written as $\nabla \cdot \mathcal{F}^n_{IMEX}$.
Taking conservative moments of this equation, we have,
\begin{equation}
\frac{W^{n+1}-W^{n}}{\Delta t}+\nabla\cdot\mathcal{F}^n = 0 . \label{eq:MacroEquation}
\end{equation}
So the conservative variables $W^{n+1}$ can be obtained explicitly.
Then, $g^{n+1}$ is fully determined because there is one to one correspondence between macroscopic variables and the equilibrium state.
Therefore, we have
\begin{equation}
f^{n+1} = \frac{\tau^{n+1}}{\tau^{n+1}+\Delta t}[f^n-\Delta t \nabla\cdot \mathcal{F}^n_{IMEX}] + \frac{\Delta t}{\tau^{n+1}+\Delta t} g^{n+1}.
\end{equation}
Assuming $f^n = g^n + O(\tau)$, then reformulate $f^{n+1}$ as following,
\begin{equation}
f^{n+1} = g^{n+1}-\tau^{n+1}[\frac{f^{n+1}-f^n}{\Delta t}+\nabla\cdot \mathcal{F}^n_{IMEX}].
\end{equation}
The solution $f^{n+1}$ is
\begin{equation}
f^{n+1} = g^{n+1}-\tau^{n+1}[\frac{g^{n+1}-g^n}{\Delta t}+\mathbf{u}\cdot\nabla g^n] + O(\tau^2).
\end{equation}
It also gives the following estimation,
\begin{equation}
f^{n+1} = g^{n+1}-\tau^{n+1}[g^{n+1}_t+\mathbf{u}\cdot\nabla g^{n+1}] + O(\tau\Delta t) + O(\tau^2).
\end{equation}

It shows that the dissipation at $n+1$ time step is originally determined by the spatial derivative at $n$ time step.
This kind of dissipation is constrained by a factor $\tau$. Therefore, the dissipation induced by this term is of order $O(\tau \Delta t)$ in the above equation.
However, as shown below, the majority of the physical dissipation comes from the convection term.
As we know, the concepts of viscosity and heat conductivity are based on the continuum assumption.
The dissipation of a scheme in the continuum limit is determined by the macroscopic equations, such as the equation (\ref{eq:MacroEquation}).
Only through the flux term in Eq.(\ref{eq:MacroEquation}) can a distribution function effect the dissipation of the scheme.
Therefore, it is insufficient to get insight of the dissipation by merely making comparison between $f^{n+1}$ and the Chapman-Enskog expansion.
So in order to understand and estimate the real dissipation in IMEX AP scheme, we must investigate the discretization of the convection term in Eq.(\ref{eq:MacroEquation})
which will be fully discussed later.


\subsection{Unified gas kinetic scheme}
\label{sec:IntrUGKS}
The key ingredient of the unified gas kinetic scheme is to couple the collision effect into the convection term.
The earliest application of this ideal is the gas kinetic scheme (GKS) for the Navier-Stokes solutions \cite{Xu93phd,Xu2001}.
The unified gas kinetic scheme (UGKS) makes a major step forward \cite{ugks1,ugks1_1,ugks2,ugks3}.
It solves the kinetic equation in all flow regimes including the free molecular and the Navier-Stokes limits.
Formally, the UGKS can be written as,
\begin{eqnarray}
\frac{f^{n+1}-f^{n}}{\Delta t}+\nabla\cdot\mathcal{F}^n_{UGKS} = \frac{1}{2}[\frac{g^{n+1}-f^{n+1}}{\tau^{n+1}}+\frac{g^{n}-f^{n}}{\tau^{n}}] .
\label{eq:ugks1}
\end{eqnarray}
The main difference between the UGKS and the IMEX AP scheme is the discretization of convection terms. The detailed construction is introduced later.

Explicitly solving $f^{n+1}$ in UGKS, we get,
\begin{eqnarray}
f^{n+1}&=&\frac{\Delta t}{2\tau^{n+1}+\Delta t}g^{n+1}+\frac{\tau^{n+1}}{\tau^n}\frac{\Delta t}{2\tau^{n+1}+\Delta t}g^n
 +\frac{2\tau^n-\Delta t}{2\tau^{n+1}+\Delta t}f^{n} \nonumber\\
 &&-\frac{2\tau^{n+1}}{2\tau^{n+1}+\Delta t}\nabla\cdot\mathcal{F}^n_{UGKS}\Delta t ,
 \label{eq:ugks2}
\end{eqnarray}
which is
\begin{eqnarray}
f^{n+1} &=& g^{n+1}-\tau^{n+1}[\frac{f^{n+1}-f^{n}}{\Delta t}+\nabla\cdot\mathcal{F}^n_{UGKS}] \nonumber\\ &&+\tau^{n+1}[\frac{g^{n}-f^{n}}{\tau^n}-(\frac{f^{n+1}-f^{n}}{\Delta t}+\nabla\cdot\mathcal{F}^n_{UGKS})] .
\label{eq:ugks1}
\end{eqnarray}
Then, with the estimation for the original BGK model approximation,
\begin{eqnarray}
\frac{g^{n}-f^{n}}{\tau^n}-(\frac{f^{n+1}-f^{n}}{\Delta t}+\nabla\cdot\mathcal{F}^n_{UGKS}) = O(\Delta t),
\end{eqnarray}
we obtain,
\begin{eqnarray}
f^{n+1} &=& g^{n+1}-\tau^{n+1}[g_t^{n+1}+\mathbf{u}\cdot\nabla g^{n+1}]+O(\tau \Delta t) + O(\tau^2).
\end{eqnarray}
The above estimation is identical with IMEX AP scheme.
It is suggested that the collision terms always induce an additional dissipation term proportional to $O(\tau \Delta t)$.
It is a very good estimation for the Navier-Stokes AP limit. But the numerical discretization of the convection terms, such as $\mathcal{F}^n_{IMEX}$ and $\mathcal{F}^n_{UGKS}$,
deviate the IMEX and UGKS schemes, which are shown below.

\subsection{Discretization of convection terms}
\subsubsection{IMEX AP Scheme}
Taking one dimensional case as an example, the discrete scheme is written in a conservative form,
and the numerical flux at $x_{i+\frac{1}{2}}$ is given below,
\begin{eqnarray}
\mathcal{F}_{IMEX} = uf_{i+1/2}.
\end{eqnarray}
Then, the cell interface distribution function can be defined as,
\begin{eqnarray}
f_{i+1/2}^l = f_i+\frac{\Delta x}{2} S^l, \nonumber\\
f_{i+1/2}^r = f_{i+1}-\frac{\Delta x}{2} S^r, \\
f_{i+1/2} = H(u) f_{i+1/2}^l+(1-H(u)) f_{i+1/2}^r ,  \nonumber
\end{eqnarray}
where $H(u)$ is Heaviside function,
\begin{equation}
  H(u) = \left\{\begin{array}{l@{\quad}l}
    0, & u < 0, \\
    1, & u \geq 0 ,\end{array}\right.
\end{equation}
and $S^l,S^r$ represent slopes with a slope limiter. In this study, the van Leer slope limiter is used.

\subsubsection{UGKS}
The numerical flux of UGKS is derived by considering the local analytical solution,
\begin{eqnarray}
\Delta t \mathcal{F}_{UGKS} &=& \int_0^{\Delta t} {u} f(t) dt  . \nonumber
\end{eqnarray}
The distribution function used to evaluate the interface flux is given below,
\begin{eqnarray}
    f({x},{u},t) &=& e^{-t/\tau}f_0({x}-{u}t)  \nonumber \\
        && + \frac{1}{\tau}\int_0^{t}g({x'},{u},t')
        e^{-(t-t')/\tau}dt', \label{eq:localsolution}
\end{eqnarray}
where $ {x'} =  {x}-{u}(t-t')$.
Here $f_0$ is constructed almost the same as $f_{i+1/2}$ in IMEX AP scheme,
but it is only taken as the initial condition at the beginning of each time step.
Suppose the cell interface is located at the origin $(x=0)$, the initial gas distribution function is,
\begin{eqnarray}
f_0(x) &=& f_{i+1/2}(x) = H(u)f_{i+1/2}^l(x) + (1-H(u))f_{i+1/2}^r(x) \\
&=& H(u)(f_i + (x+\frac{\Delta x}{2}) S^l) + (1-H(u))(f_{i+1} + (x-\frac{\Delta x}{2}) S^r) . \nonumber
\end{eqnarray}
The inclusion of the equilibrium part, $g({x},t,u)$, is the key of UGKS.
It is formally written as,
\begin{eqnarray}
g({x},t,u) = g_0(1+ax+At),
\end{eqnarray}
where $g_0$ is defined by the moments of $f_{i+1/2}$, namely,
\begin{eqnarray}
\int \psi g_0 du = \int \psi f_{i+1/2} du.
\end{eqnarray}
In the equilibrium expansion, $a$ and $A$ \cite{Xu2001} represent the spatial and temporal derivatives of the equilibrium state respectively.
Then the interface flux of UGKS is written as,
\begin{eqnarray}
\Delta t \mathcal{F}_{UGKS} &=& \int_0^{\Delta t} u f(t) dt \nonumber \\
 &=& u\{\tau(1-e^{-\Delta t/\tau})\ (H(u)f_{i+1/2}^l+(1-H(u))f_{i+1/2}^r) \nonumber\\
 && +\tau(\tau(e^{-\Delta t/\tau}-1)+\Delta t e^{-\Delta t/\tau})\ u(H(u)S^l+(1-H(u))S^r)  \nonumber \\
 && +(\Delta t - \tau(1-e^{-\Delta t/\tau}))\ g_0 \nonumber\\
 && + \tau(2\tau(1-e^{-\Delta t/\tau})-\Delta t(1+e^{-\Delta t/\tau}))\ au g_0 \nonumber\\
 && +(\frac{1}{2}(\Delta t)^2+\tau(\tau(1-e^{-\Delta t/\tau})-\Delta t))\ A g_0\} .\label{eq:UGKSFlux}  \nonumber
\end{eqnarray}
Although this expression seems very complicated, it could be written in a concise form with clear physical meaning. This will be discussed later.



\subsection{Boundary condition}
Boundary condition is a crucial part of a numerical scheme. We need a unified solid boundary condition to complete AP scheme. Before proposing such a boundary condition, we introduce the kinetic fully diffusion boundary condition (KBC) at first.
Suppose normal direction towards the solid is in the positive direction. The incoming distribution function at boundary
is represented by $f_{gas}$, which is interpolated from inner cells. The distribution function of the particles emitted from the
solid wall is a Maxwellian distribution function,
\begin{equation}
  g_{W} = \rho_W\left(\frac{1}{2\pi RT_W}\right)^{\frac{3}{2}}
    e^{-\frac{1}{2RT_W}((\mathbf{u}-\mathbf{U}_W)^2)},
\end{equation}
where $T_W$ and $\mathbf{U}_W$ are given by solid boundary.
Then no penetration condition is used to determine $\rho_W$, i.e.,
\begin{equation}
  \int_{u-U_W \geq 0}
    (u-U_W)f_{gas} d\mathbf{u} =
   \int_{u-U_W < 0}
    (u-U_W)g_W d\mathbf{u}.
\end{equation}
The above Maxwell boundary condition works very well in the rarefied flow regime, but
it may not work properly in the continuum flow limit.
In the above Maxwell boundary construction, the Knudsen layer will be enlarged
since it assumes that the particles coming from gas collide with the solid boundary directly without suffering from any inter-molecule collisions.
If we change the Maxwell boundary condition to a simple bounce back boundary condition for the non-slip boundary condition, it cannot
work properly for the rarefied gas flow where velocity slip appears.
Therefore, in order to get a valid boundary condition for both rarefied and continuum flows, a multiscale boundary condition (MBC) is required
for all Knudsen number regime.
Here, we propose a multiscale boundary condition. The ideal is to use local solution again, namely, $f_{gas}$ is replaced by formula (\ref{eq:localsolution}).
Similar boundary condition has been developed for the radiative transfer problem \cite{Mieussens2013}.
The gas initial spatial distribution is reconstructed as usual. The local equilibrium state $g_0$ is determined from the solid boundary condition, say,
\begin{eqnarray}
T_0 = T_W , \nonumber\\
U_0 = U_W . \nonumber
\end{eqnarray}
The pressure of $g_0$ is set to the same pressure at the adjacent cell center.
Then, the density of $g_0$ is fully determined,
\begin{eqnarray}
\rho_0 = \rho_{gas} T_{gas} / T_W  . \nonumber
\end{eqnarray}
This boundary condition only changes the incoming distribution function.
The particles emitted from the boundary also obey the fully diffusion assumption.
In the following section, both the original Maxwell kinetic boundary condition (KBC) and the current multiscale boundary condition (MBC)
are implemented in the numerical schemes.

\section{Numerical experiments}
\begin{figure}[h]
    \parbox[t]{0.48\textwidth}{
    \includegraphics[totalheight=6cm, bb = 90 35 690 575, clip =
    true]{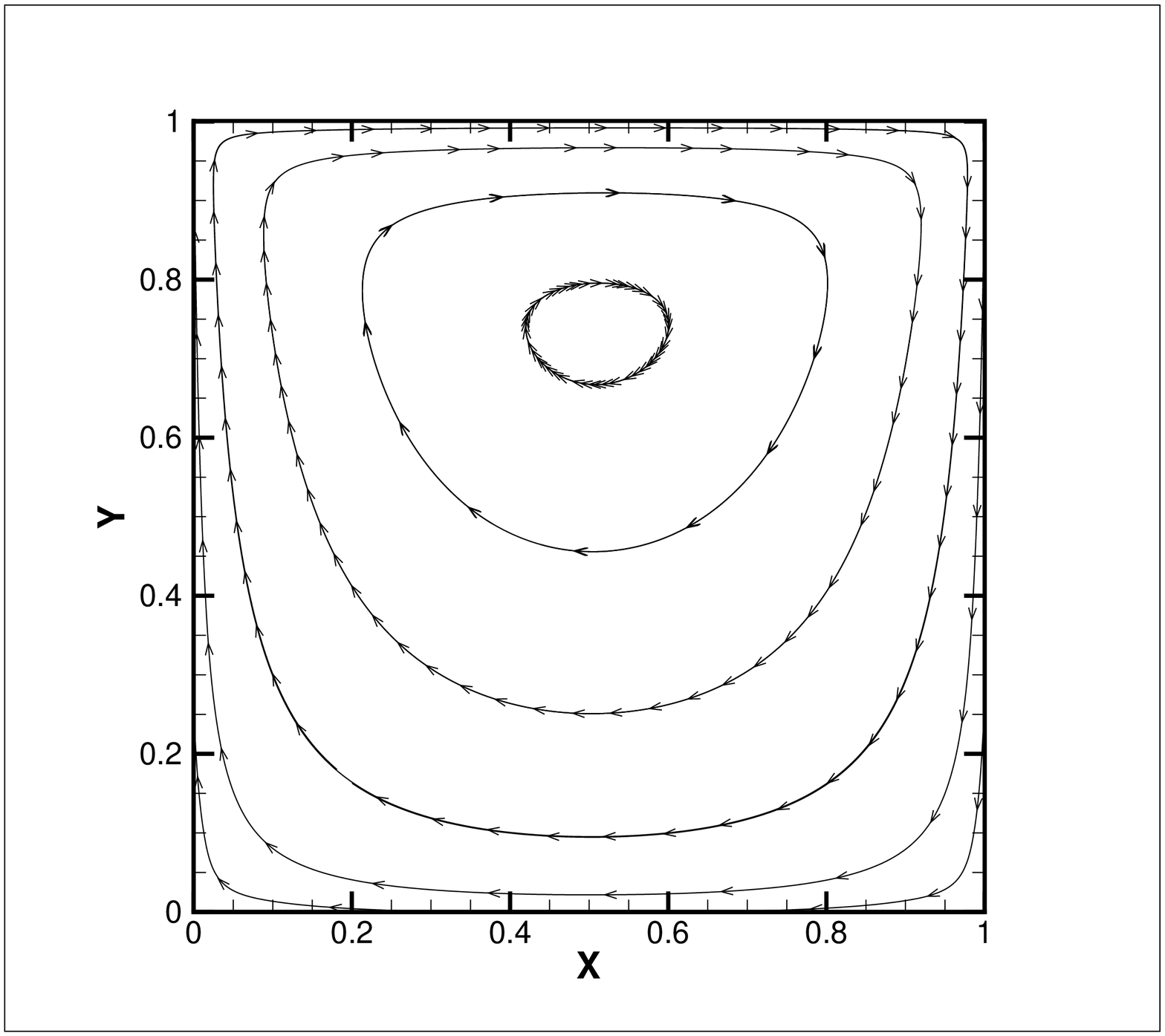}
    }
    \hfill
    \parbox[t]{0.48\textwidth}{
    \includegraphics[totalheight=6cm, bb = 90 35 690 575, clip =
    true]{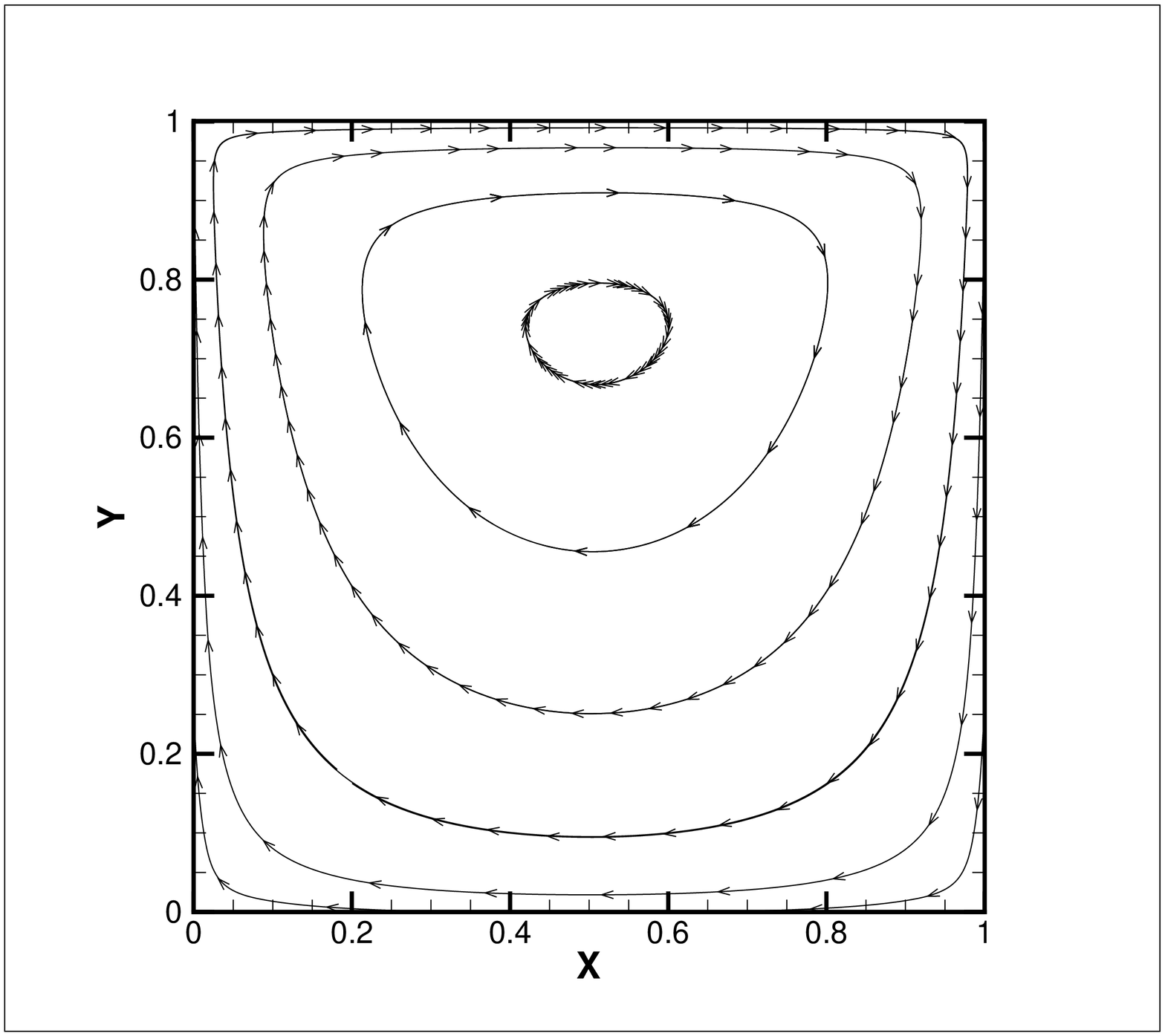}
    }
    \caption{The stream lines for lid-driven cavity flow at $\mbox{Kn} = 0.1$ with
    $61\times 61$ mesh points in the physical space and $40 \times 40$ mesh points in the velocity space.
    The left is from IMEX AP scheme and the right is from UGKS.}
    \label{fig:Kn01FlowField}
\end{figure}

\begin{figure}[h]
    \parbox[t]{0.48\textwidth}{
    \includegraphics[totalheight=6cm, bb = 90 35 690 575, clip =
    true]{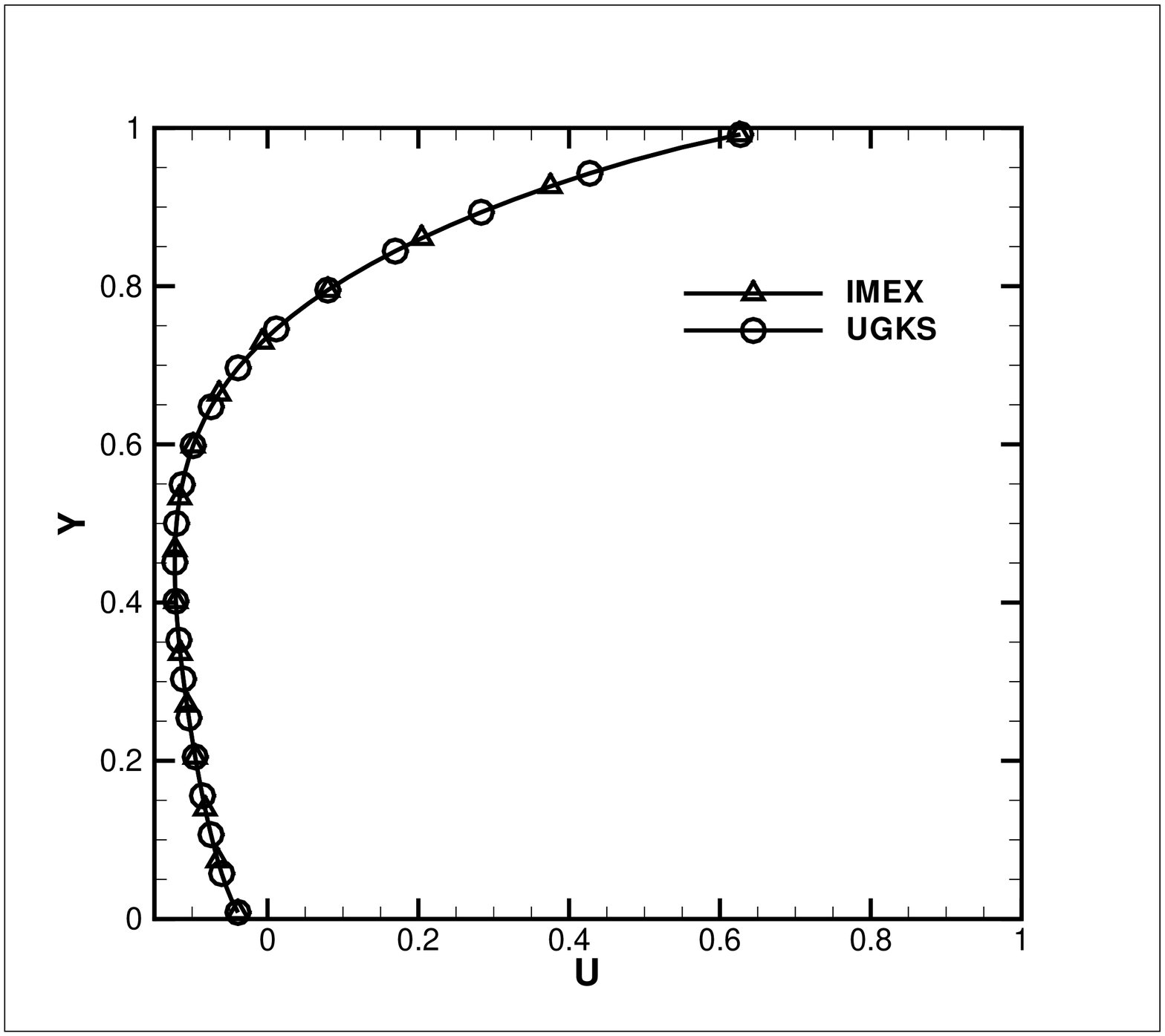}
    }
    \hfill
    \parbox[t]{0.48\textwidth}{
    \includegraphics[totalheight=6cm, bb = 90 35 690 575, clip =
    true]{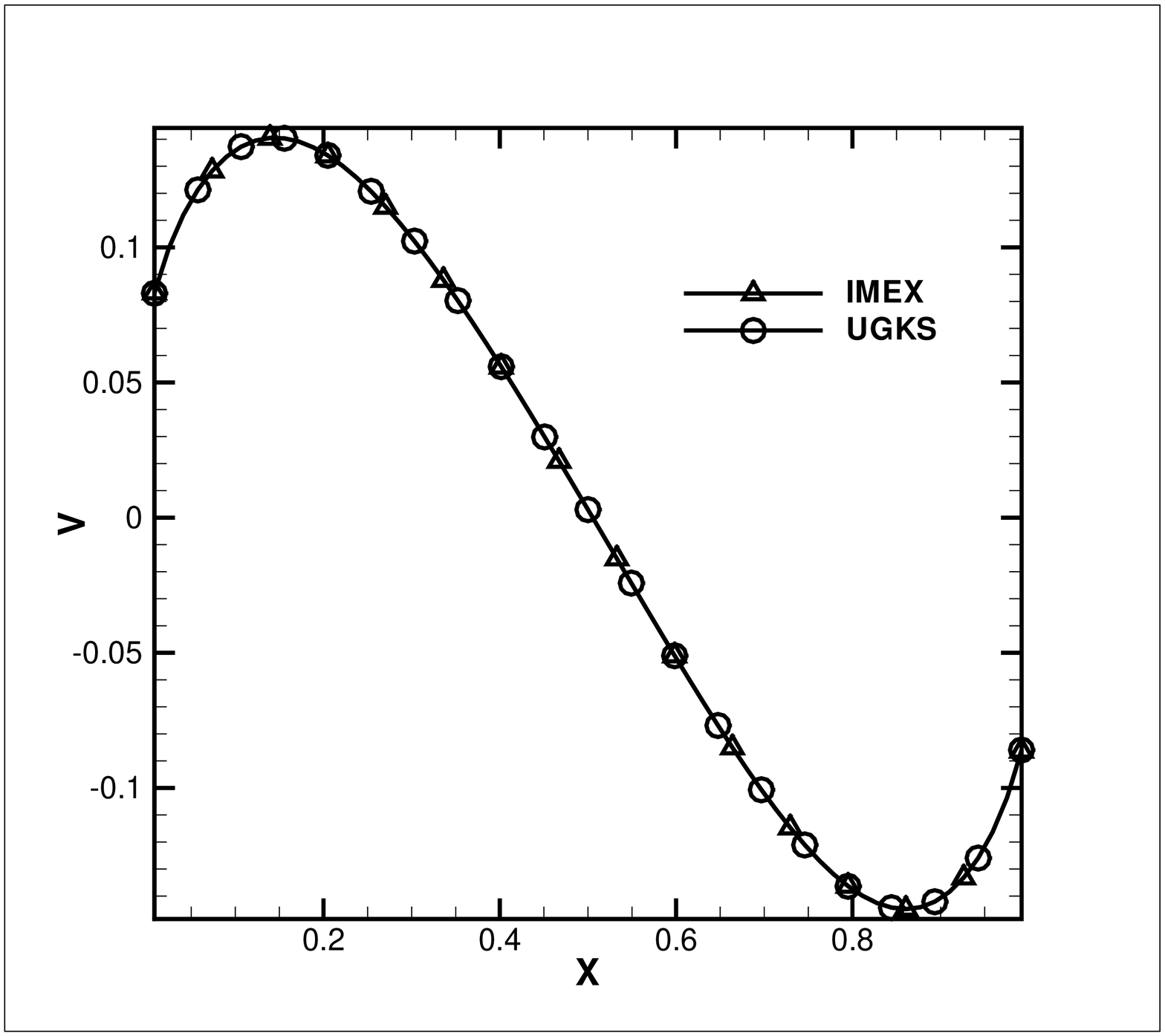}
    }
    \caption{The velocity profiles along the vertical and horizontal central lines at $\mbox{Kn} = 0.1$.}
    \label{fig:Kn01lines}
\end{figure}

The Navier-Stokes limit is characterized by shear flow.
This kind of flow feature does not appear in one dimensional case.
Thus, we choose lid-driven cavity flow as a benchmark test case.
For the first calculation, we consider the transition flow regime to make sure the performance of two schemes for the capturing rarefied flow behavior.
The working gas is argon with temperature~viscosity relation $\mu\sim T^{\omega}$, where $\omega = 0.81$. The reference viscosity is defined as
\begin{eqnarray}
\mu_{ref} = \frac{30}{(7-2\omega)(5-2\omega)}\frac{\rho\lambda\sqrt{2\pi RT}}{4} ,
\label{eq:refvis}
\end{eqnarray}
where $\lambda$ is the mean free path. We choose $\mbox{Kn} = 0.1$ for the first case. The Mach number is about $0.15$. The top wall moves from left to right with a unit velocity.
The computational domain is $[0,1]\times[0,1]$ which is covered by $61\times 61$ uniform mesh points in the physical space.
The discretization in velocity space is $40\times 40$. The ratio between the time step and the relaxation time, $\Delta t/\tau$, is about $0.014$. Figure \ref{fig:Kn01FlowField} shows the flow field from two schemes.
They are almost identical. The velocity profiles (fig. \ref{fig:Kn01lines}) from both schemes are indistinguishable along the vertical central line and the horizontal central line.
Therefore, it is confirmed that both UGKS and IMEX AP scheme obtain the same result in transition flow regime.
Both KBC and MBC are tested, and they perform equally well.

\begin{figure}[h]
    \parbox[t]{0.48\textwidth}{
    \includegraphics[totalheight=6cm, bb = 90 35 690 575, clip =
    true]{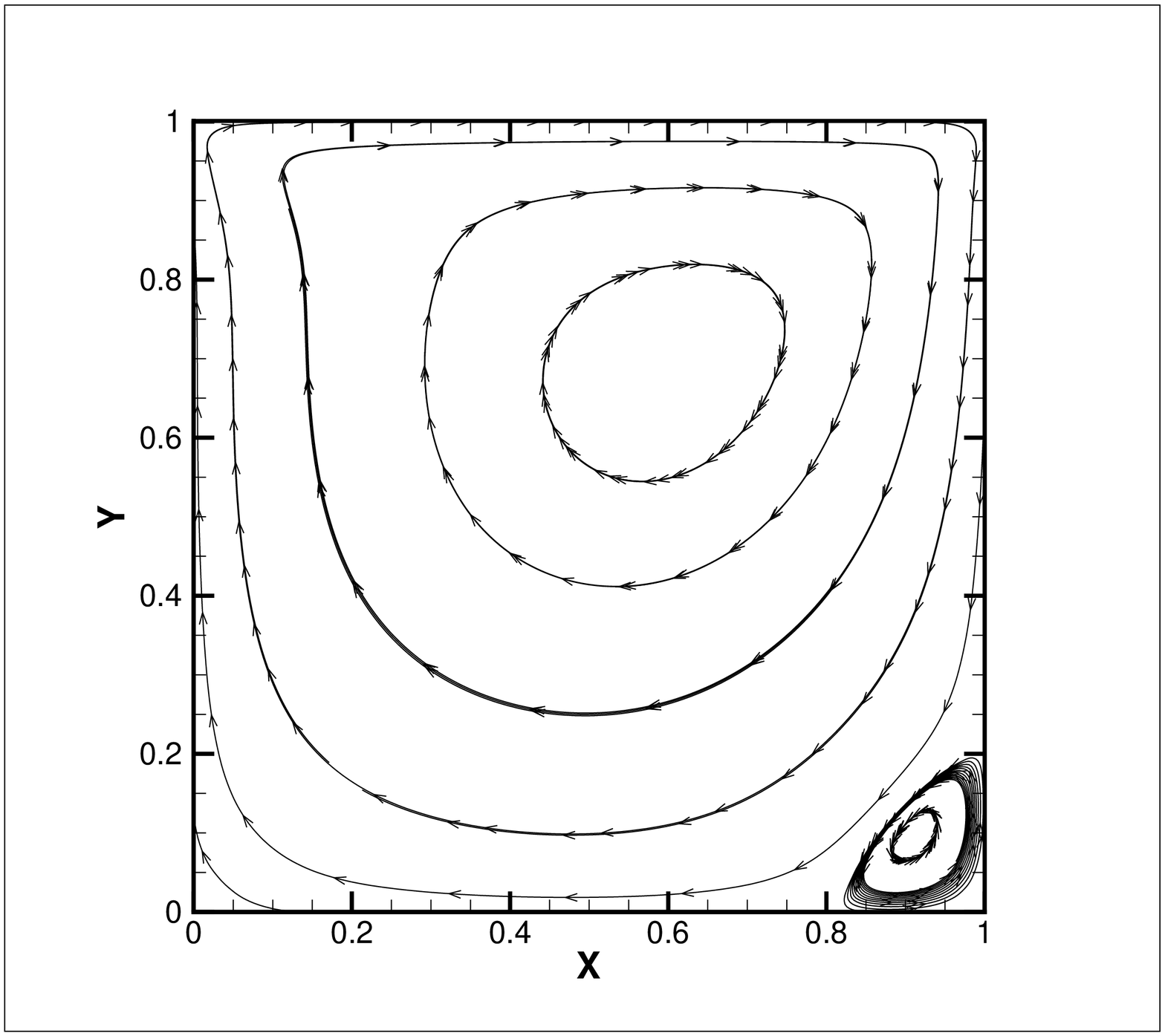}
    }
    \hfill
    \parbox[t]{0.48\textwidth}{
    \includegraphics[totalheight=6cm, bb = 90 35 690 575, clip =
    true]{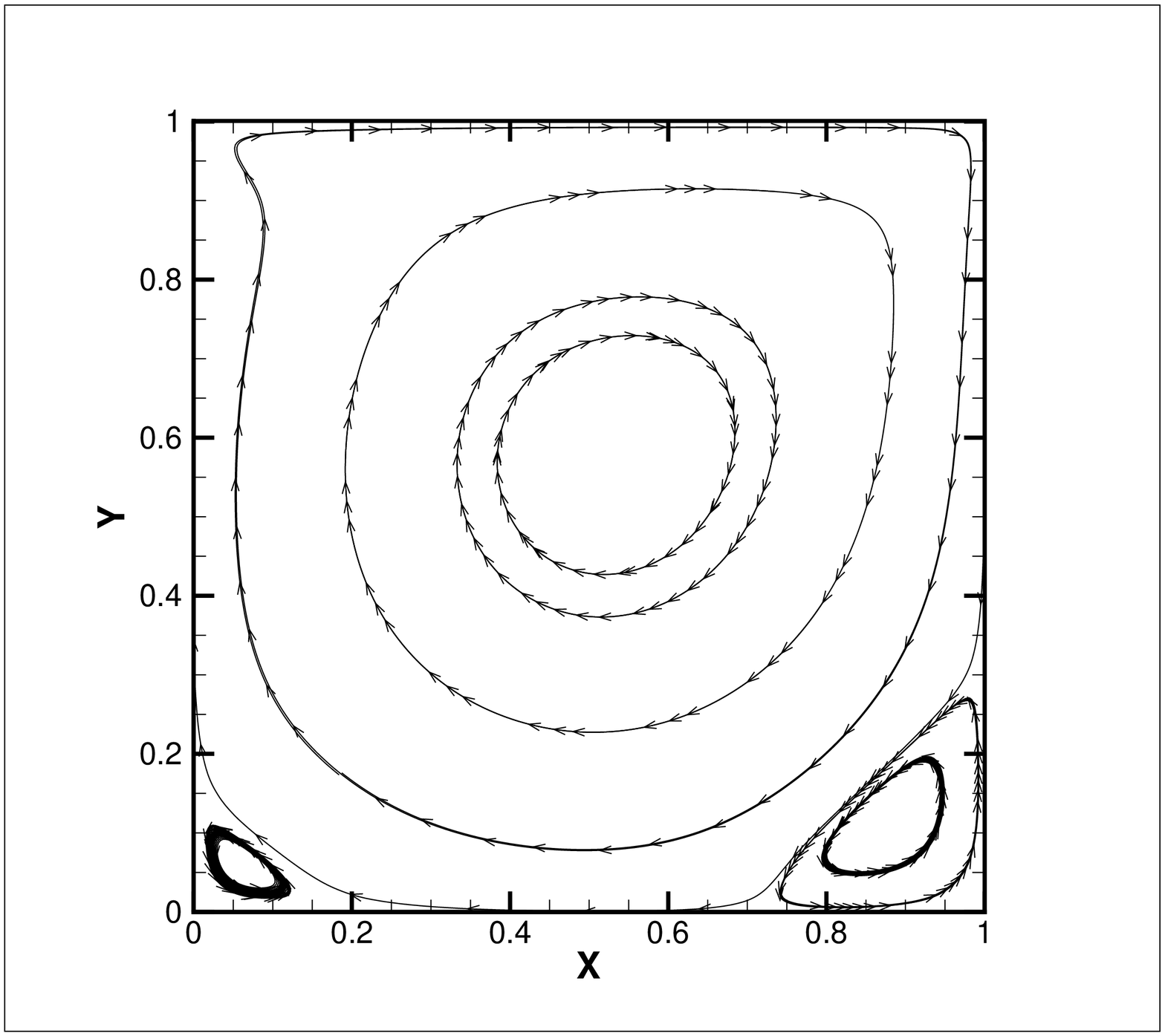}
    }
    \caption{The stream lines for lid-driven cavity flow at $\mbox{Re} = 1000$. The left is from IMEX AP scheme, and right is from UGKS.}
    \label{fig:Re1kFlowField}
\end{figure}

\begin{figure}[h]
    \parbox[t]{0.48\textwidth}{
    \includegraphics[totalheight=6cm, bb = 90 35 690 575, clip =
    true]{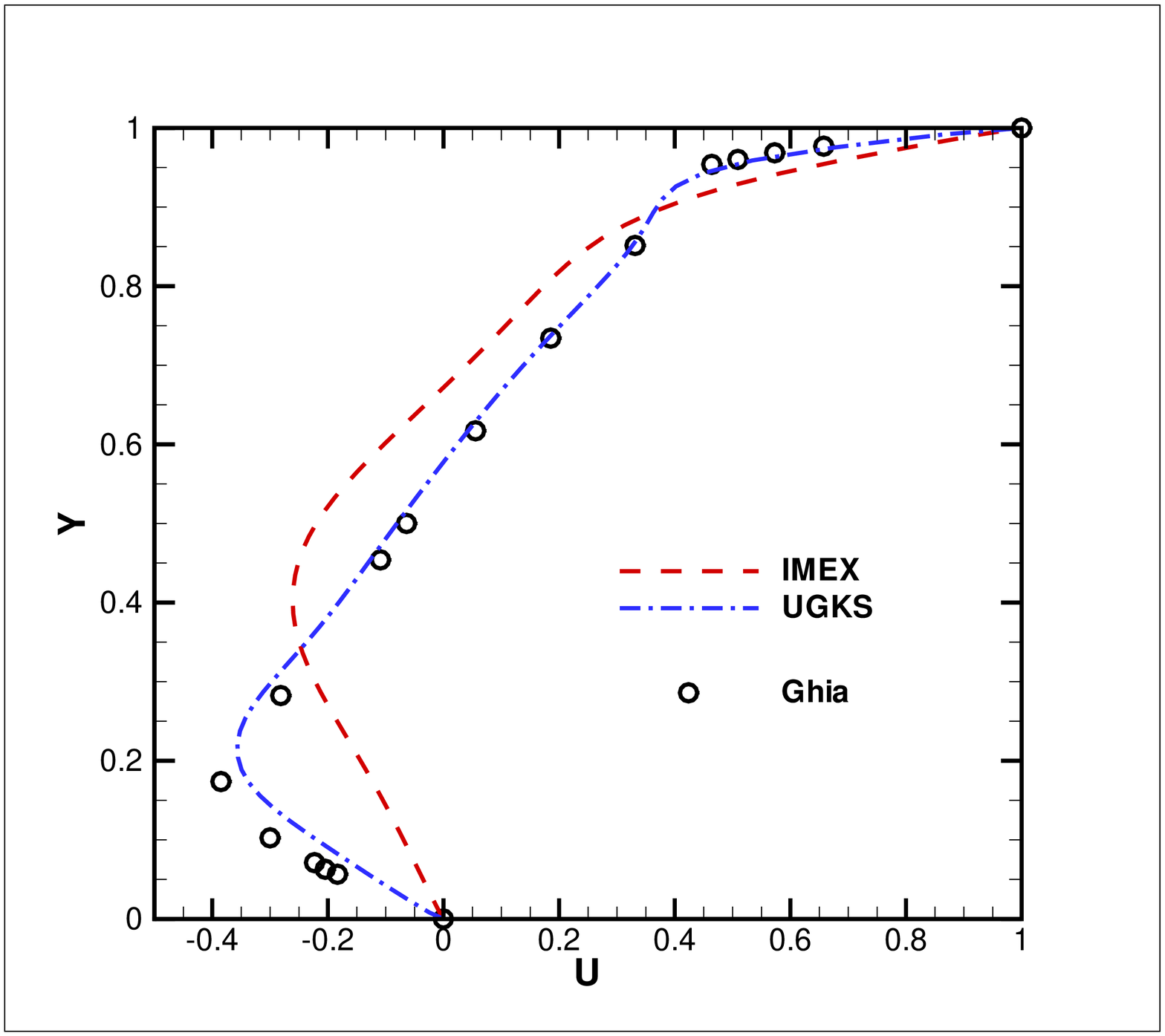}
    }
    \hfill
    \parbox[t]{0.48\textwidth}{
    \includegraphics[totalheight=6cm, bb = 90 35 690 575, clip =
    true]{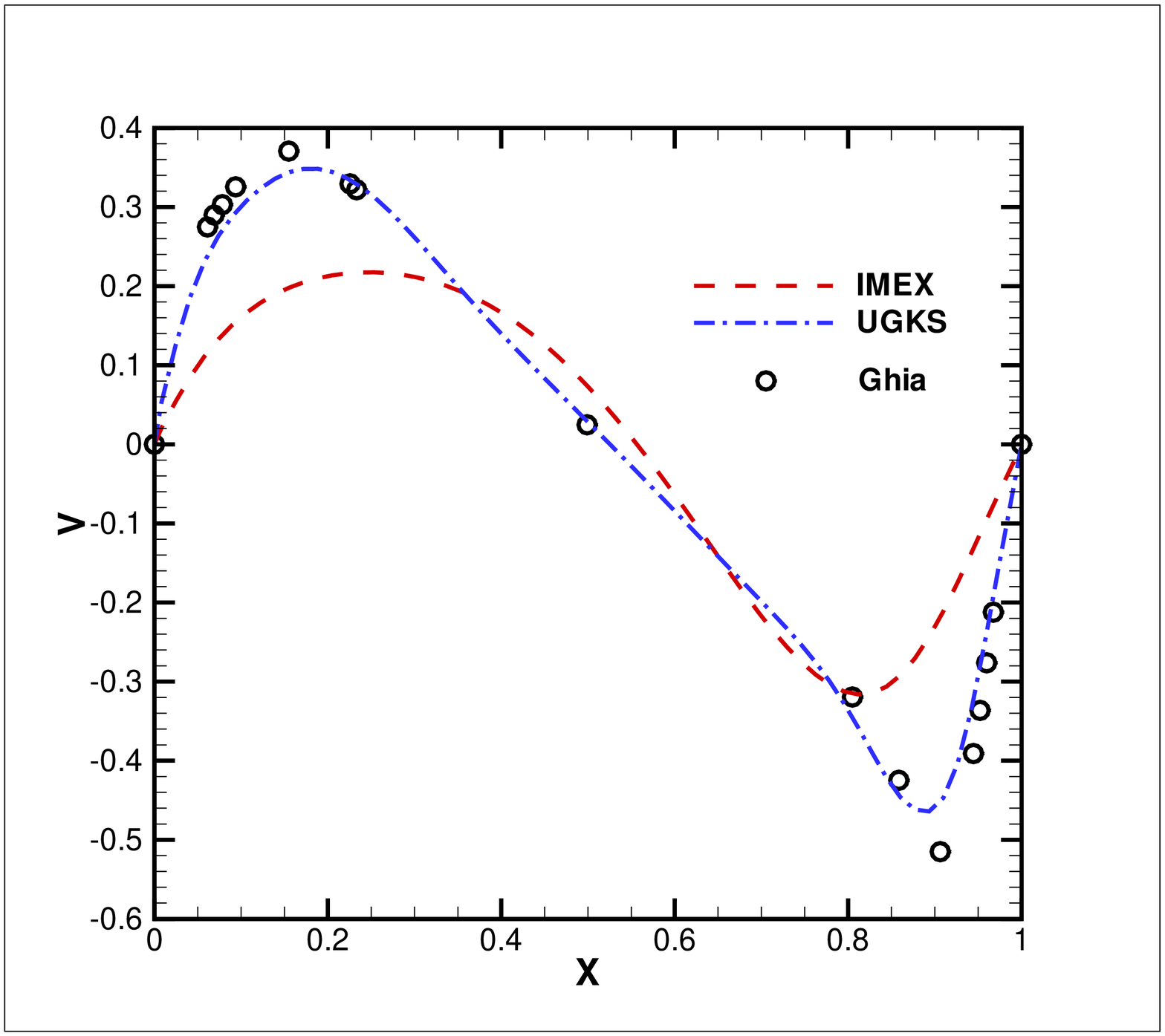}
    }
    \caption{The velocity profiles along the vertical and horizontal central lines at $\mbox{Re} = 1000$.}
    \label{fig:Re1klines} 
\end{figure}

Then, we reduce the Knudsen number until the Reynolds number being equal to $1000$. Correspondingly, $\Delta t/\tau$ increases to about $7.2$.
The mesh in physical space keeps $61\times 61$, and the mesh in velocity space is $20\times 20$.
Note that as small Knudsen number the mesh points in the velocity space can be reduced due to the regularity of the distribution function.
The case of $\mbox{Re} = 1000$ is in a typical continuum flow regime.
The numerical results are compared with classical benchmark solutions from Ghia \cite{Ghia1982}. Figure \ref{fig:Re1kFlowField}
shows the flow field at $\mbox{Re} = 1000$.
The boundary condition is MBC for UGKS, and KBC for IMEX AP scheme.
Compared with the solution in transition regime, the eddies are developed at the bottom corners.
The vortex from UGKS is obviously more intensive than that from IMEX AP scheme.
The velocity profiles along the central line shown in figure \ref{fig:Re1klines}  demonstrate that the UGKS get the results closer to the benchmark solutions.

\begin{figure}[h]
\centering
    \parbox[t]{0.48\textwidth}{
    \includegraphics[totalheight=6cm, bb = 90 35 690 575, clip =
    true]{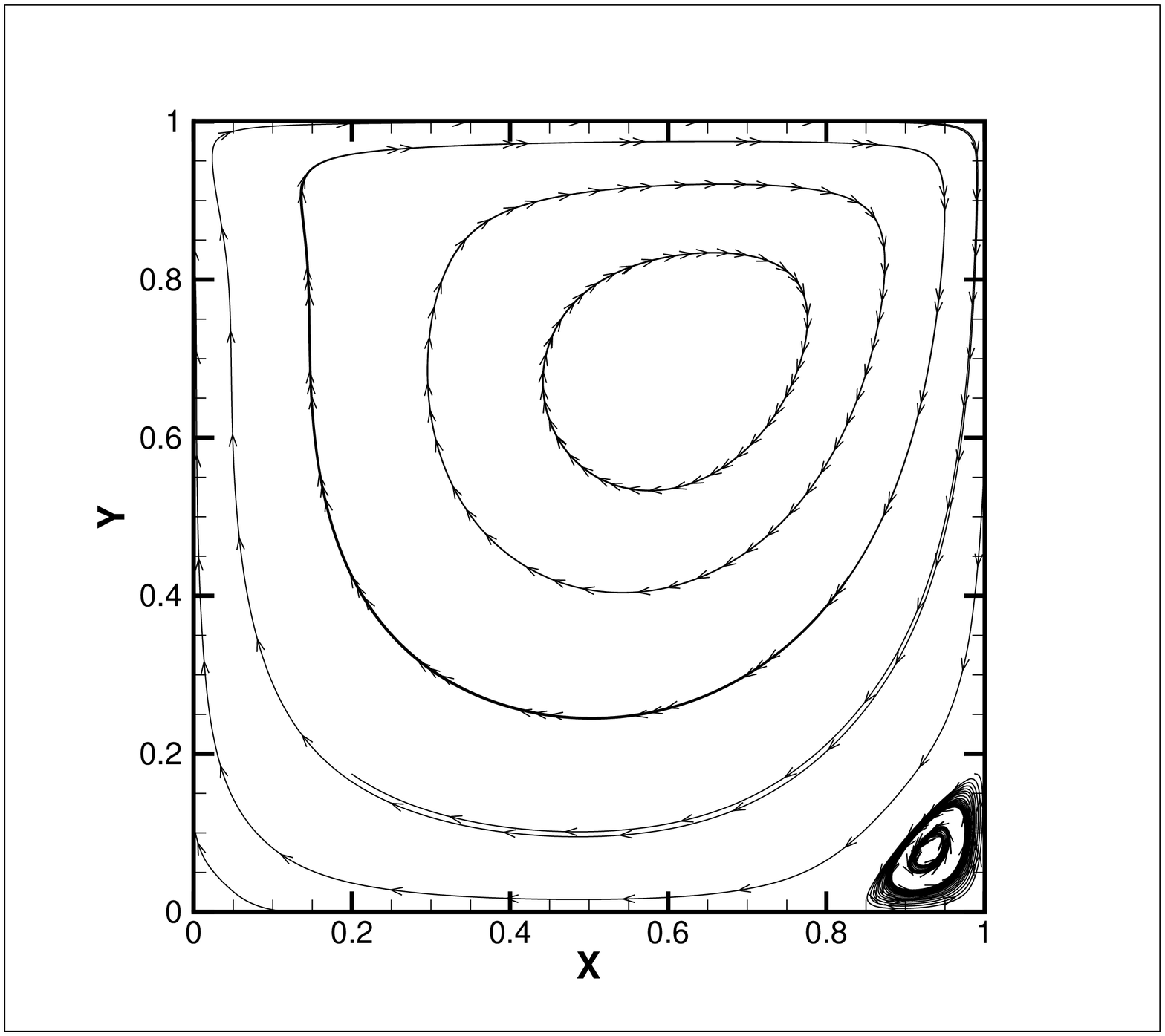}
    }
    \parbox[t]{0.48\textwidth}{
    \includegraphics[totalheight=6cm, bb = 90 35 690 575, clip =
    true]{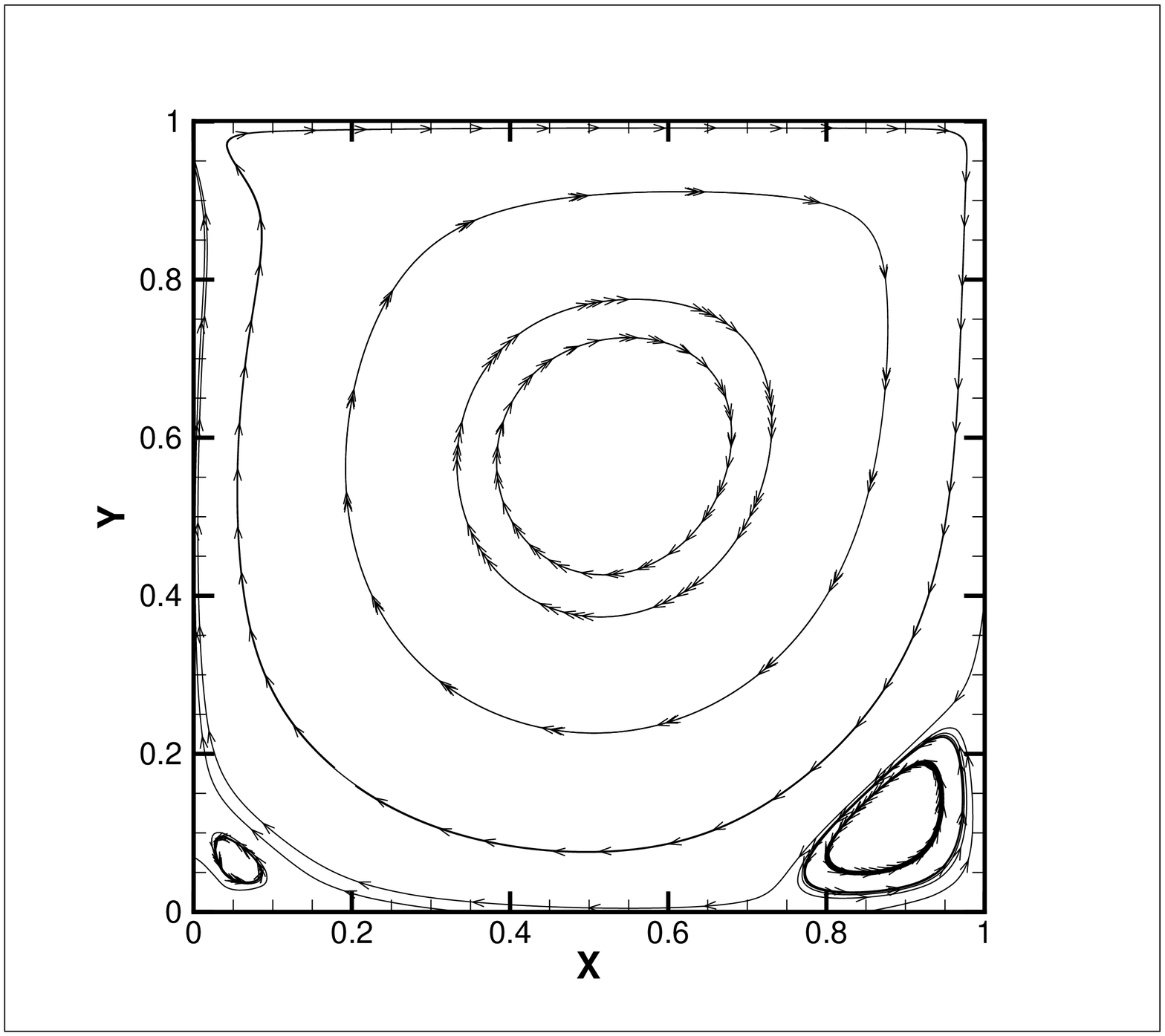}
    }
    \caption{The stream lines derived by AP schemes with different boundary conditions at $\mbox{Re} = 1000$. The left is IMEX+MBC, and the right is UGKS+KBC.}
    \label{fig:KBCMBCFlowField} 
\end{figure}

\begin{figure}[h]
    \parbox[t]{0.48\textwidth}{
    \includegraphics[totalheight=6cm, bb = 90 35 690 575, clip =
    true]{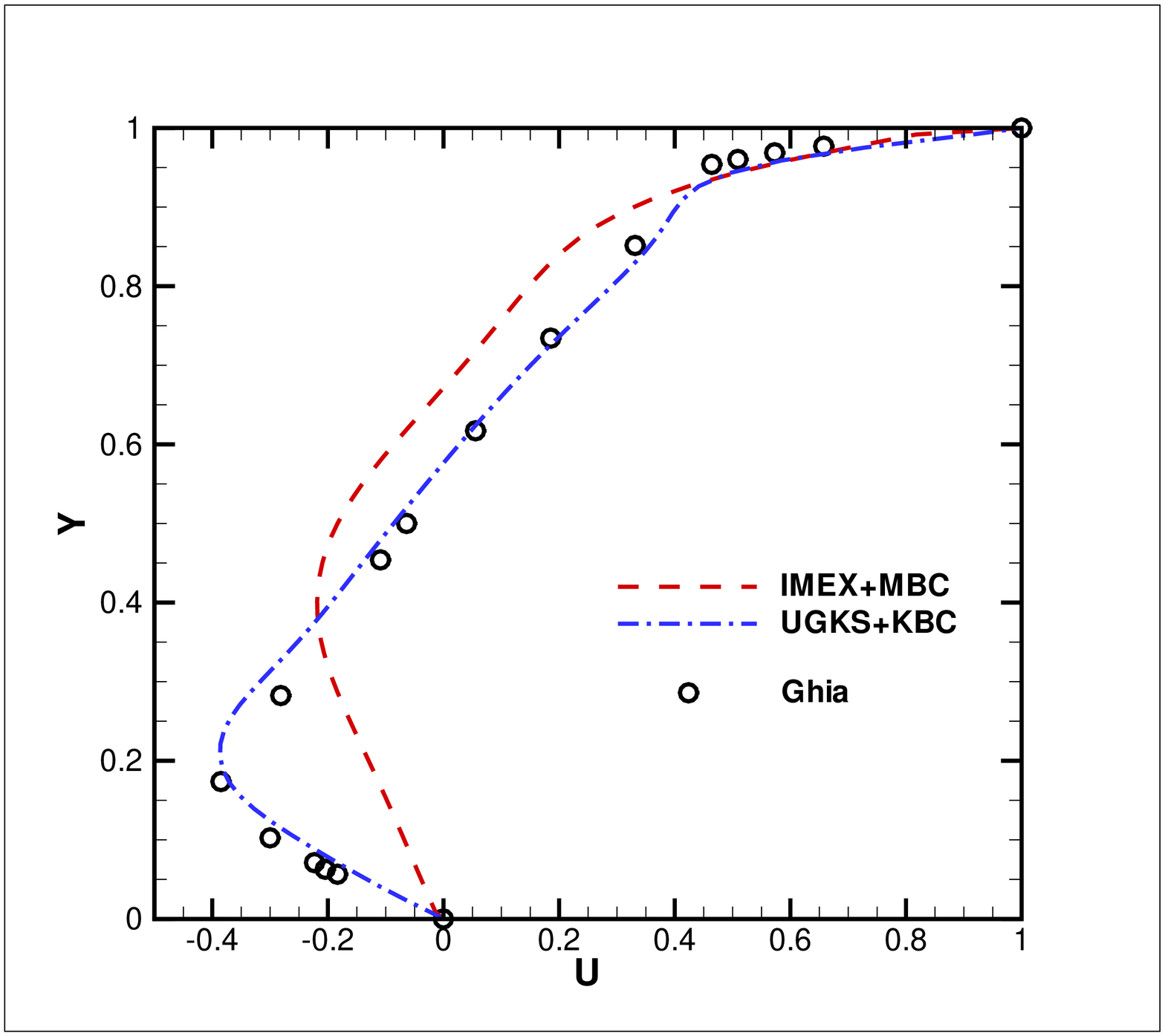}
    }
    \hfill
    \parbox[t]{0.48\textwidth}{
    \includegraphics[totalheight=6cm, bb = 90 35 690 575, clip =
    true]{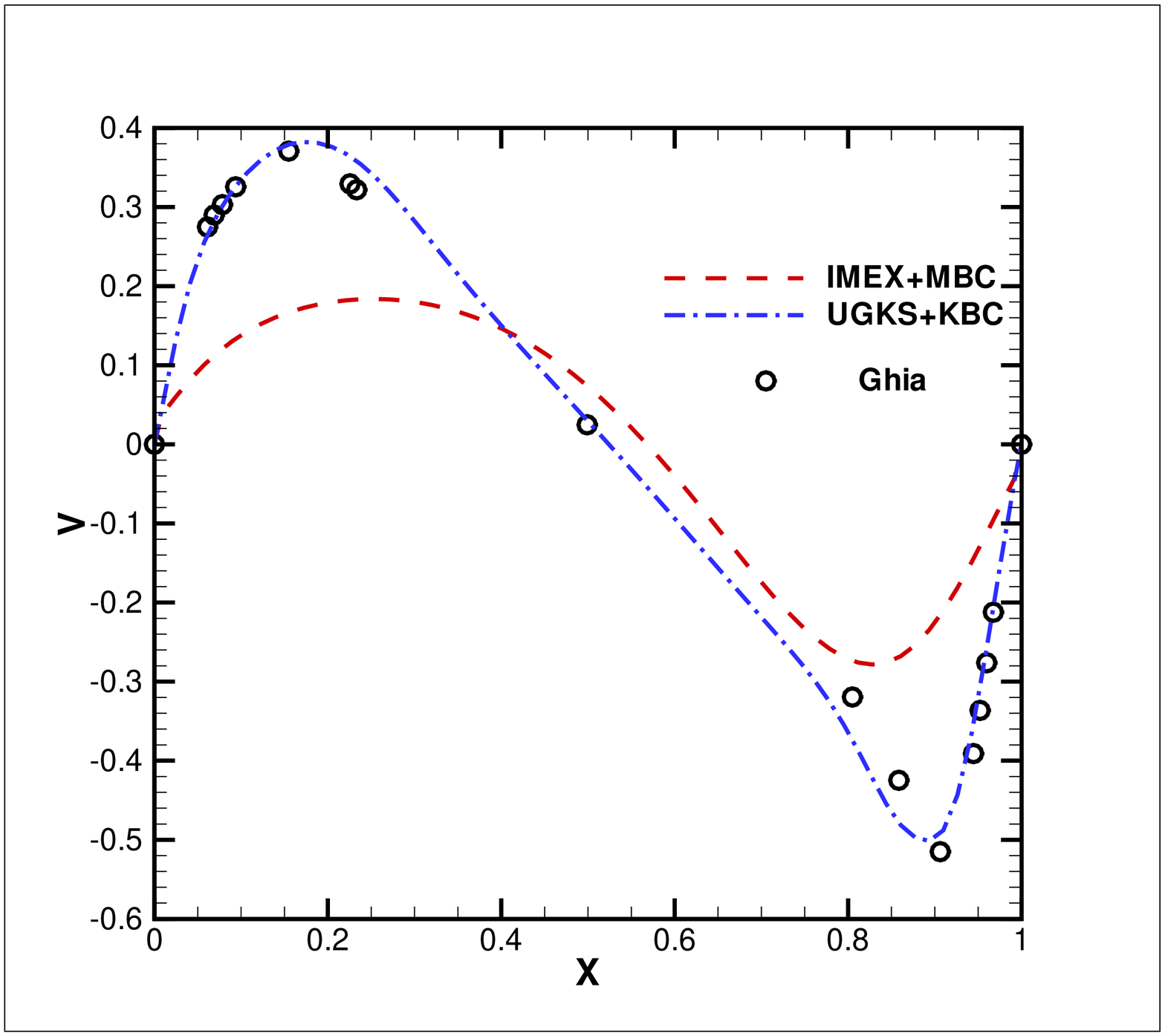}
    }
    \caption{The comparison of IMEX AP and UGKS using the same kinetic fully diffusion boundary condition.}
    \label{fig:KBCMBClines} 
\end{figure}

The boundary condition could strongly affect the final results. Therefore, in order to have a complete picture
we design another two combinations of schemes and boundary conditions, namely,
UGKS$+$KBC and IMEX$+$MBC. The MBC in the IMEX AP scheme uses UGKS solver for the flux at boundary, since IMEX AP doesn't have  the corresponding equilibrium part in the numerical flux.
As shown in figure \ref{fig:KBCMBCFlowField}, both results seem more dissipative than UGKS+MBC.
The eddies in the corners are also smaller. However, the maximum vertical velocity (see fig. \ref{fig:KBCMBClines}) from UGKS+KBC is much greater than the one derived by UGKS+MBC,
since the KBC enlarges the Knudsen layer, and thus enhances the momentum exchange between gas and solid wall.
An enlarged Knudsen layer smears the eddies near the boundary.
Therefore, the results from UGKS+KBC preserve higher velocity in interior flow region and smaller vortexes in the corners.
On the other hand, the IMEX+MBC doesn't change too much in comparison with IMEX+KBC.
It seems that the IMEX AP solver could not recover the hydrodynamic NS solution due to its inaccurate boundary layer capturing.
According to the simulation results, the UGSK+MBC is the best scheme among four combinations.

\begin{figure}[h]
    \parbox[t]{0.48\textwidth}{
    \includegraphics[totalheight=6cm, bb = 90 35 690 575, clip =
    true]{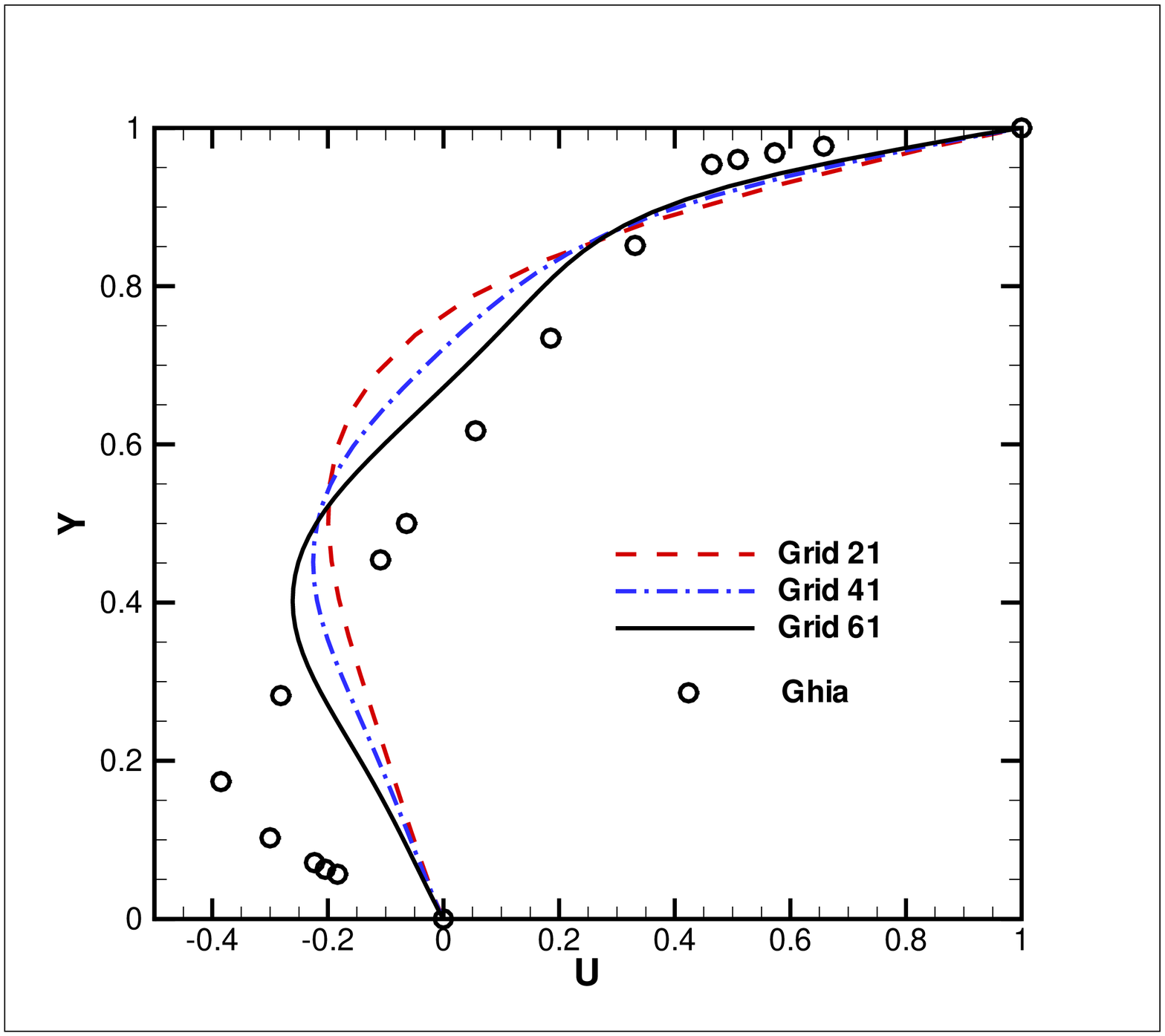}
    }
    \hfill
    \parbox[t]{0.48\textwidth}{
    \includegraphics[totalheight=6cm, bb = 90 35 690 575, clip =
    true]{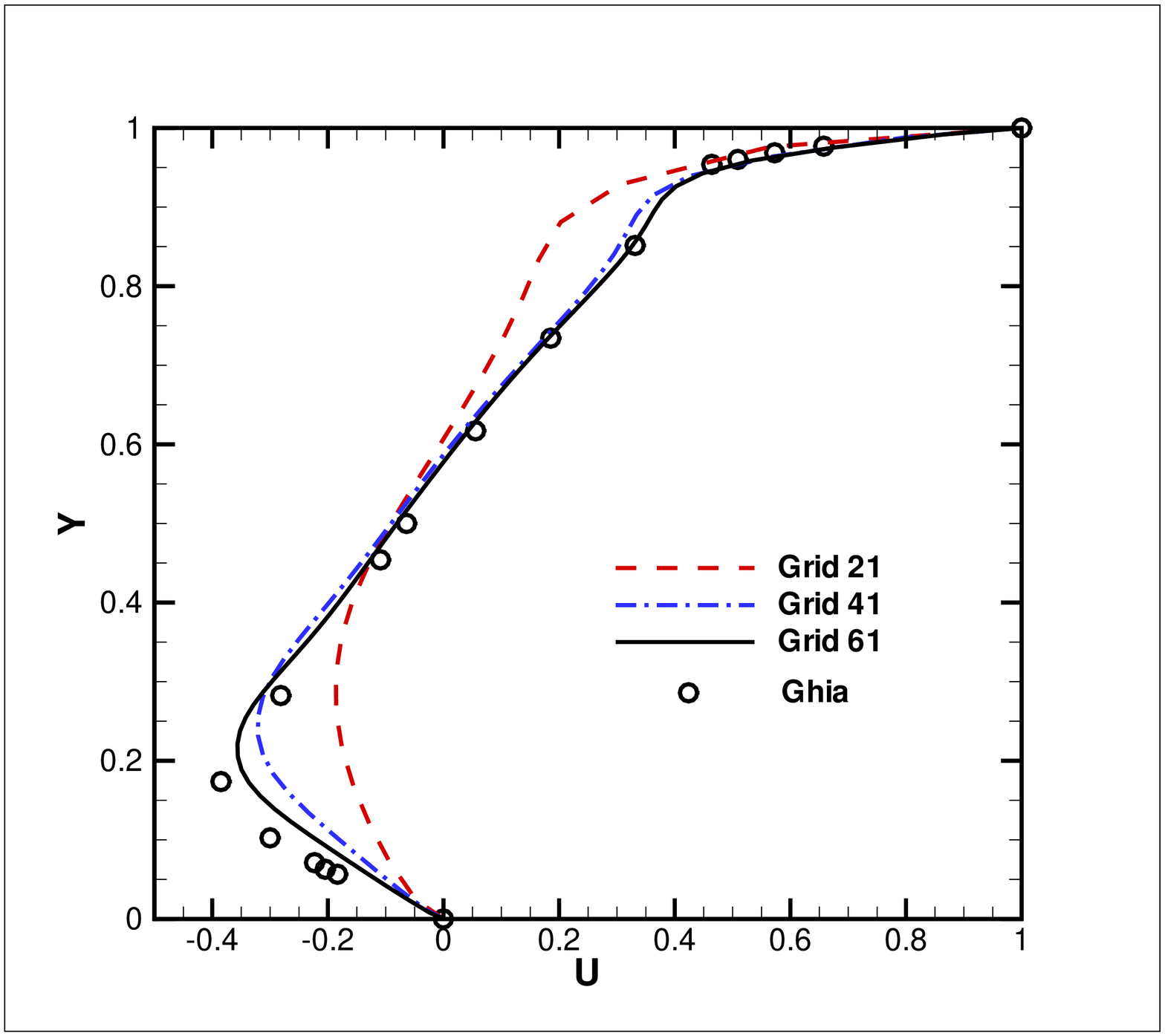}
    }
    \caption{The U velocity along the central line $x=0.5$ on different mesh points in the physical space  of $(21 \times 21 , 41 \times 41, 61 \times 61)$.
    The left is from IMEX AP, and the right is from UGKS.}
    \label{fig:3casex} 
\end{figure}

\begin{figure}[h]
    \parbox[t]{0.48\textwidth}{
    \includegraphics[totalheight=6cm, bb = 90 35 690 575, clip =
    true]{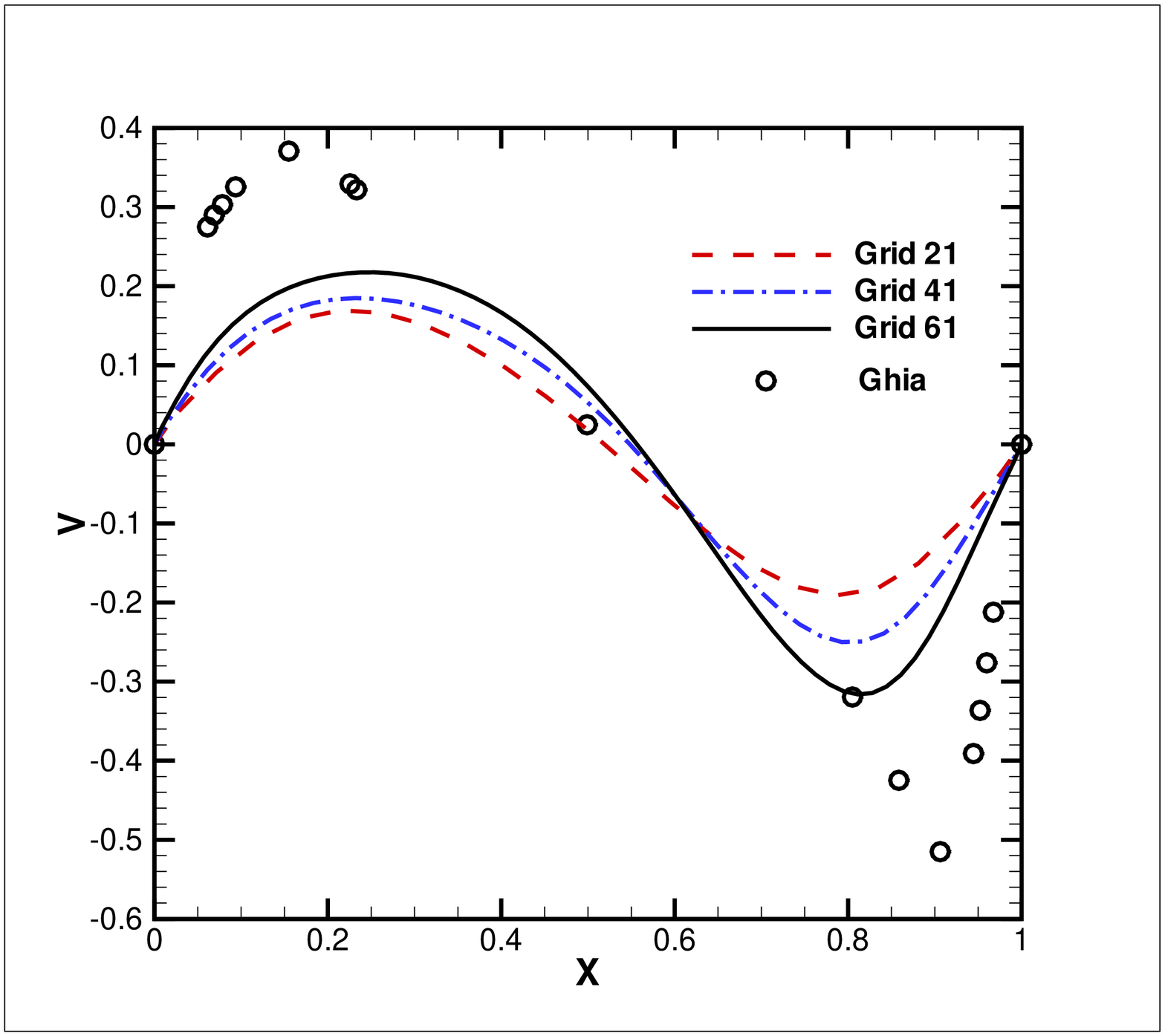}
    }
    \hfill
    \parbox[t]{0.48\textwidth}{
    \includegraphics[totalheight=6cm, bb = 90 35 690 575, clip =
    true]{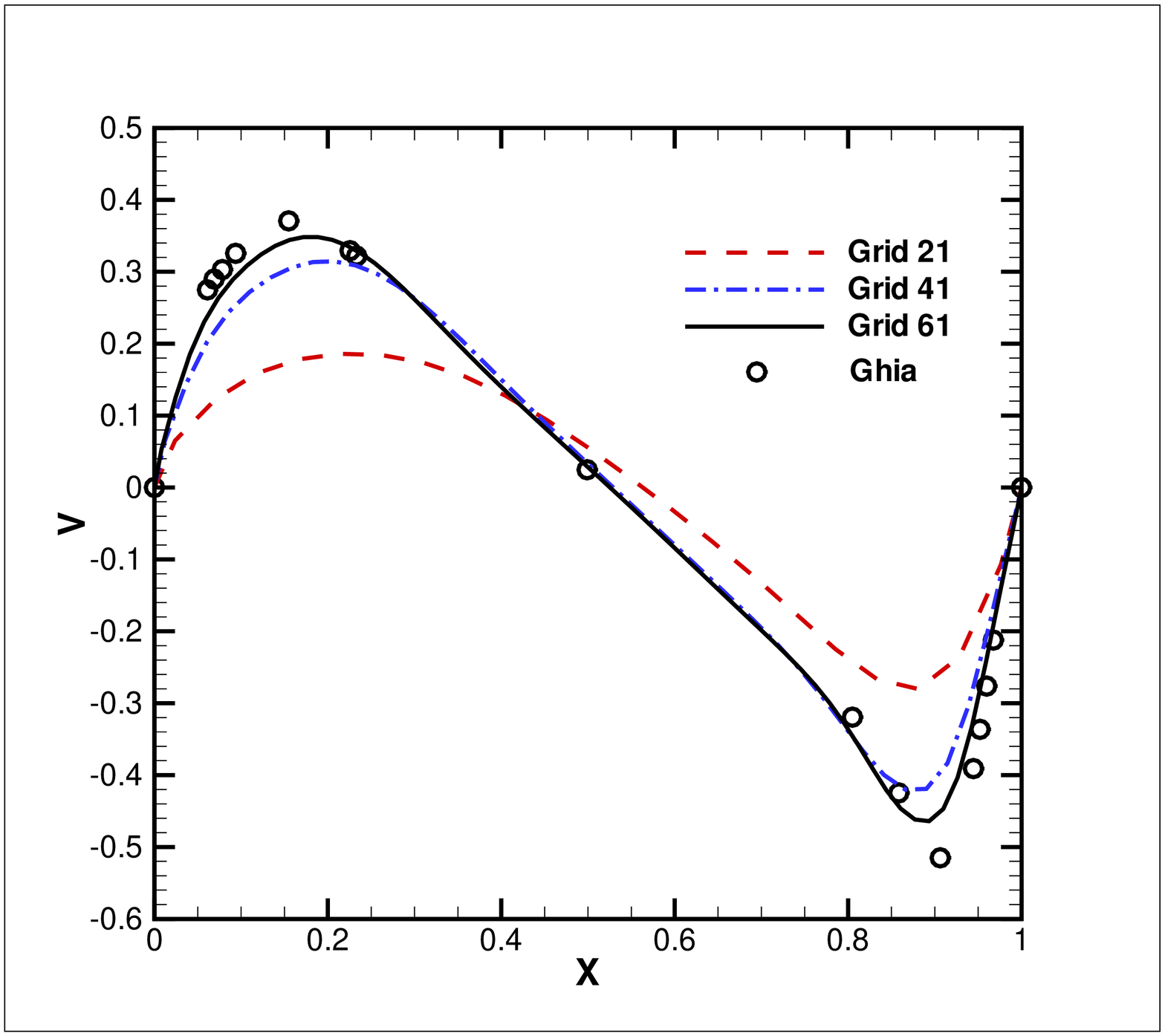}
    }
    \caption{The V velocity along the central line $y=0.5$ on different grids. The left is from IMEX AP, and the right is from UGKS.}
    \label{fig:3casey} 
\end{figure}

As a further test, we study the performance of UGKS+MBC and IMEX+KBC on coarse meshes.
As shown in figure \ref{fig:3casex} and \ref{fig:3casey}, the UGKS gets better velocity profiles on every mesh, which is not so sensitive to the mesh points.
Even on the $21\times 21$ mesh points, the UGKS still get the two obvious eddies at the bottom corners, which is qualitatively correct.

These numerical results clearly show that the UGKS preserves AP property in the Navier-Stokes limit.
We attribute such property to the analytical solution used in both flux solver and boundary condition of UGKS,
because the analytical solution reflects the coupling of convection and collision. In continuum flow regime, these two processes are inseparable.

\section{Analysis about Dissipation from Convection Terms}

\subsection{IMEX}
As shown in section \ref{sec:APschemes}, the deviation from the Chapman-Enskog expansion due to collision term treatment
is only $O(\tau \Delta t)$. In fact, we have evaluated the collision terms in both IMEX and UGKS schemes introduced in section \ref{sec:IntrIMEX} and \ref{sec:IntrUGKS}.
The results are indistinguishable. In other words, it confirms that the small differences in the collision process don't deteriorate simulation results at all.
Furthermore, a distribution function could effect the dissipation mainly through the flux term in Eq.(\ref{eq:MacroEquation}).
As shown by the numerical experiments, although the implicit discretization of collision term guarantees the small deviation from the local equilibrium,
the discretization of convection terms could still generate large artificial dissipation.
Therefore, the main dissipation in AP schemes in the continuum limit comes from the convection term, specifically, $<\mathbf{u}\cdot\nabla f>$.
For example, the kinetic flux vector splitting (KFVS) scheme generates large numerical dissipation which is proportional to time step
even with exactly the local equilibrium initial condition.
The IMEX scheme shares the same streaming mechanism as KFVS in the convection process.
The numerical flux of IMEX scheme can be reformulated as,
\begin{eqnarray}
\mathcal{F}_{IMEX} = uf_{i+1/2}=u(g_0 + [H(u) f_{i+1/2}^l+(1-H(u)) f_{i+1/2}^r - g_0]), \label{eq:IMEXFlux2}
\end{eqnarray}
where $g_0$ is the Maxwellian distribution corresponding to the discontinuous distribution function $f_{i+1/2}$.
The terms inside square bracket are the dissipation terms.

The distribution function, $f_{i+1/2}$, is composed of the contributions from two different sides.
It is a non-equilibrium state unless these two parts are coming from a uniform flow.
In other words, this kind of distribution function deviates from equilibrium state and generates dissipation.
Considering the asymptotic process, we assume the two parts of $f_{i+1/2}$ are interpolated from local Chapman-Enskog distributions.
\begin{eqnarray}
f_{i+1/2}^{l} &=& f_i+\frac{\Delta x}{2} S^l = g_i+\frac{\Delta x}{2} \partial_x g_i + O(\tau)  \\
f_{i+1/2}^{r} &=& f_{i+1} - \frac{\Delta x}{2} S^r = g_{i+1}-\frac{\Delta x}{2} \partial_x g_{i+1} + O(\tau) .
\end{eqnarray}
With a small variation inside each cell, it is reasonable to assume that $g+\frac{\Delta x}{2} \partial_x g$ presents a local Maxwellian, namely,
$f_{i+1/2}^{l} = g^l + O(\tau)$ and $f_{i+1/2}^{r} = g^r + O(\tau)$, where $g^l$ and $g^r$ are independent of $\tau$.
Given $\rho^l,U^l,T^l$ represents the macroscopic variables corresponding to $f_{i+1/2}^{l}$, the counterpart for $f_{i+1/2}^{r}$ is represented by $\rho^l+\rho',U^l+U' ,T^l+T'$.
Define the viscous term as below,
\begin{eqnarray}
p_{1} = \int_{-\infty}^{+\infty} (u-U)(u-U)f_{i+1/2} du - \rho_{i+1/2} R T_{i+1/2}.
\end{eqnarray}
Obviously,
$$\frac{\partial p_1}{\partial \rho'} \neq 0,\ \ \frac{\partial p_1}{\partial U'} \neq 0,\ \ \frac{\partial p_1}{\partial T'} \neq 0.$$
Since $g^l$ and $g^r$ are independent of $\tau$,
we can get
$$p_1 = O(\rho')+O(U')+O(T')+O(\tau).$$



As the van Leer limiter gets first order reconstruction at extreme point, we only consider the case where $f$ is a monotonic function.
Consider a continuous reconstruction, the van Leer limiter reads,
\begin{equation}
S^l = \frac{2S_{i-1}S_i}{S_{i-1}+S_i},\ \ S^r = \frac{2S_{i}S_{i+1}}{S_{i}+S_{i+1}},
\end{equation}
where $S_{i} = (f_{i+1}-f_{i})/\Delta x$. Therefore,
\begin{eqnarray}
f^{l}_{i+1/2}-f^{r}_{i+1/2} &=& f_i-f_{i+1}+S_i[\frac{f_i-f_{i-1}}{S_{i-1}S_i}+\frac{f_{i+2}-f_{i+1}}{S_{i+1}S_i}] \nonumber\\
&=& (f_i-f_{i+1})[1-(\frac{f_i-f_{i-1}}{f_{i+1}-f_{i-1}}+\frac{f_{i+2}-f_{i+1}}{f_{i+2}-f_i})] \nonumber\\
&=& (\partial_x f_{i+1/2} \Delta x +O(\Delta x^3))[\frac{\partial_x^2 f_{i}\Delta x}{2\partial_x  f_{i}}-\frac{\partial_x^2 f_{i+1}\Delta x}{2\partial_x  f_{i+1}}+O(\Delta x^3)] \nonumber\\
&\sim & O(\Delta x^2) .
\end{eqnarray}
As a result,
\begin{eqnarray}
p_1 \sim \rho' \sim U' \sim T' \sim <f^{l}_{i+1/2}-f^{r}_{i+1/2}> \sim O(\Delta x^2).
\end{eqnarray}
In this sense, the IMEX AP scheme with the van Leer slope limiter presents a dissipation proportional to $\Delta x$ at extreme points and $(\Delta x)^2$ in other continuous region when the flow system approaches to the continuum limit.

\subsection{UGKS}
The numerical flux of UGKS (Eq.(\ref{eq:UGKSFlux})) is derived by considering a local analytical solution.
The expression is very complicated. However, these coefficients related with $\Delta t$ and $\tau$ adjust the weights of different spatial discretization.
For example, when Knudsen number goes to infinity, $f_0$ constructed through slope limiter will dominate to recover the collisionless limit.
Meanwhile, in this limit the hydrodynamic part $g_0$, which is nearly constructed by central difference, is totally ignored.
On the other hand, when Knudsen number approaches to zero, namely, $\Delta t/\tau \rightarrow \infty$, the numerical flux only deviates from the Maxwellian flux a little bit,
\begin{eqnarray}
\mathcal{F}_{UGKS}
 &=& u \{g_0[1 - \tau (au + A) + \frac{1}{2}\Delta t A] \nonumber \\
 && + \frac{\tau}{\Delta t}[H(u)f_i+(1-H(u))f_{i+1} - g_0]+O(\tau^2)\}. \label{eq:UGKSFlux1}
\end{eqnarray}
The second term of the UGKS flux is identical with the deviation part in the formula (\ref{eq:IMEXFlux2}).
But, this term, generating a dissipation proportional to the square of cell size as IMEX does, is suppressed  by a factor $\frac{\tau}{\Delta t}$
and gets its contribution less than $O(\tau)$. However, the forepart of UGKS flux is identical with the Chapman-Enskog expansion
expressed in the Lax-Wendroff scheme \cite{Ohwada2004}. The leading order of the deviation from the Maxwellian distribution is $O(\tau)$.
Formula (\ref{eq:IMEXFlux2}) and Formula (\ref{eq:UGKSFlux1}) present the Euler AP and the  NS AP respectively.
In fact, in the continuum limit the collision term is so strong. No matter what kind of initial distribution function is,
the distribution function is collapsing to the local equilibrium and the flux should be the same as the Chapman-Enskog NS expansion.
The formula (\ref{eq:UGKSFlux1}) demonstrates this property, which is the most important key for NS AP schemes.
Unfortunately, many AP methods, as simple as the IMEX AP scheme introduced previously, fail to provide such a mechanism
to drive the distribution function to the local NS distribution function within the convection process.


\subsection{Coupling between Convection and Collision terms}
As shown in last subsection, the collision effect plays an important role in the convection process
in order to recover the Navier-Stokes limit.
This kind of coupling is also very important in the treatment of the collision process inside each control volume.
The numerical schemes designed for collision term target on the reproduction of the Chapman-Enskog expansion.
Even with the name of "implicit-explicit", the "implicit" approach is not a sufficient condition to recover the Chapman-Enskog NS expansion.
It can be only achieved by including the convection term into evaluation of the collision process \cite{Bennoune2008,Filbet2010,Filbet2011,Li2012}.
For example, in the splitting method, exactly solving the collision term \cite{coron1991} can only provide a distribution function
close to Euler limit \cite{Bennoune2008}. Consider a relaxation problem,
\begin{eqnarray}
\frac{\partial f}{\partial t} &=& \frac{g(t)-f(t)}{\tau} , \nonumber\\
f(0) &=& f_0.
\end{eqnarray}
The collision process is totally decoupled from the convection term. The exact solution in the above equation is
\begin{eqnarray}
f(t) &=& e^{-t/\tau}f_0 + e^{-t/\tau}\int_0^{t} \frac{g(t')}{\tau} e^{t'/\tau} dt'\nonumber \\
&=& e^{-t/\tau}f_0 + (1-e^{-t/\tau})g_0. \nonumber
\end{eqnarray}
The deviation from equilibrium state is only $O(e^{-t/\tau})$. It is so close to the equilibrium state, which can not reproduce correct dissipation of the Navier-Stokes limit.
More accurate evaluation of collision term is proposed as the exponential Runge-Kutta method \cite{Dimarco2011,Li2012}. Here, we use a simple model with constant convection to illustrate the utility of the convection term,
\begin{eqnarray}
\frac{\partial f}{\partial t} &=& \frac{g(t)-f(t)}{\tau} - \nabla\cdot \mathcal{F}^n \nonumber\\
f(0) &=& f_0.
\end{eqnarray}
The exact solution of the above equation is
\begin{eqnarray}
f(t) &=& e^{-t/\tau}f_0 + e^{-t/\tau}\int_0^{t} (\frac{g(t')}{\tau} - \nabla\cdot \mathcal{F}^n)e^{t'/\tau} dt'\nonumber \\
&=& e^{-t/\tau}(f_0-(g_0 -\tau (\partial_t g + \nabla\cdot \mathcal{F}^n))) + (g_0+\partial_t g t) - \tau (\partial_t g + \nabla\cdot \mathcal{F}^n) \nonumber \\
&=& (g_0+\partial_t g t) - \tau (\partial_t g + \nabla\cdot \mathcal{F}^n) + O(e^{-t/\tau}) .
\end{eqnarray}
It can be considered as a simplified case of formula (\ref{eq:localsolution}).
With this formulation, the Chapman-Enskog expansion are reproduced when $t/\tau \rightarrow \infty$.
The implicit treatment for collision term makes the scheme stable rather than guarantees an  AP property with respect to the Navier-Stokes limit.

Since the Chapman-Enskog expansion is a consequence of the balance between convection and collision,
the recovery of the NS solution must couple both effects. Any absence of collision or convection will not deduce the Chapman-Enskog NS expansion,
and make the scheme deviate from the Navier-Stokes limit.




\section{Criteria for Navier-Stokes Limit Asymptotic Preserving Scheme}
As implied by the Chapman-Enskog expansion, in order to capture the Navier-Stokes limit, the schemes have to be capable of capturing the term on the order $O(\tau)$.
In the past, many numerical schemes concentrate on the precise prediction of collision terms, with numerical treatment of convection term left behind.
According to the analysis and numerical results in this study, we propose three criteria for the NS AP.
To recover the Navier-Stokes limit, an AP scheme should preserve the following criteria,
\begin{itemize}
\item the numerical scheme is stable regardless of $\tau$,
\item the numerical flux is not sensitive to the initial distribution function,
\item the numerical flux is mainly contributed from a gas distribution function with the Chapman-Enskog NS expansion.
\end{itemize}
Note that the numerical flux is emphasized, not the distribution function itself.
With the same spirit, the UGKS framework has been successfully extended to the diffusion limit of linear kinetic models \cite{Mieussens2013}.

\section{Conclusion}
This paper investigates the performance of two AP schemes in the Navier-Stokes limit.
The 2D lid-driven cavity flow is used as a benchmark test case, and simulated by UGKS and IMEX AP scheme respectively.
The numerical results and analysis demonstrate that the UGKS could capture the Navier-Stokes limit more accurately.
The coupling between the convection and collision is essentially important for the Navier-Stokes limit.
The implementation of local analytical solution can reproduce such a coupling during a whole evolution time step.
The UGKS provides a general framework to construct numerical schemes in all flow regimes, because most kinetic equations have relaxation terms which can be
explicitly solved along the characteristic lines.
The setup of the benchmark test case in this paper is also important for the whole AP society, because significant amount of newly developed AP schemes have never been seriously tested
in two dimensional cases.

\section*{Acknowledgements}
This work was supported by Hong Kong Research Grant Council (621011),
and grants SRFI11SC05 and FSGRF13SC21 at HKUST.

\section*{References}

\bibliographystyle{elsarticle-num}
\bibliography{APUGKS}

\end{document}